\documentclass[%
 reprint,
nofootinbib,
 amsmath,amssymb,
 aps,
]{revtex4-2}

\usepackage{graphicx}
\usepackage{xcolor}
\usepackage{dcolumn}
\usepackage{bm}

\begin{document}

\preprint{APS/123-QED}

\title{New physics above 50 TeV: probing its phenomenology through UHECR air-shower simulations }

\author{Stylianos Romanopoulos}
 \email{sromanop@physics.uoc.gr}
\author{Vasiliki Pavliou}%
 \email{pavlidou@physics.uoc.gr}
\affiliation{%
Department of Physics \& Institute of Theoretical and Computational Physics, University of Crete, GR-70013, Heraklion, Greece\\
 Institute of Astrophysics, Foundation for Research and Technology-Hellas, Vasilika Vouton, GR-70013 Heraklion, Greece}%

\author{Theodore Tomaras}
 \email{deceased}
\affiliation{
 Department of Physics \& Institute of Theoretical and Computational Physics, University of Crete, GR-70013, Heraklion, Greece
}%

\date{\today}

\begin{abstract}
Ground based observations appear to indicate that Ultra High Energy Cosmic Rays (UHECR) of the highest energies ($>10^{18.7} \rm eV$) consist of heavy particles -- shower depth and muon production data both pointing towards this conclusion. On the other hand, cosmic-ray arrival directions at energies $>10^{18.9}{\rm \, eV}$ exhibit a dipole anisotropy, which disfavors heavy composition, since higher-Z nuclei are strongly deflected by the Galactic magnetic field, suppressing anisotropy. This is the composition problem of UHECR. One solution could be the existence of yet-unknown effects in proton interactions at center-of-mass (CM) energies $\gtrsim 50$ TeV, which would alter the interaction cross section and the multiplicity of interaction products, mimicking heavy primaries. We study the impact of such changes on cosmic-ray observables using simulations of Extensive Air-Shower (EAS), in order to  place constrains on the phenomenology of any new effects for high energy proton interactions that could be probed by $\sqrt{s}>50$ TeV collisions. We simulate showers of primaries with energies in the range $10^{17} - 10^{20} {\rm  eV}$ using the CORSIKA code, modified to implement a possible increase in cross-section and multiplicity in hadronic collisions exceeding a CM energy threshold of $50$ TeV. We study the composition-sensitive shower observables (shower depth, muons) as a function of cross-section, multiplicity, and primary energy. We find that in order to match the Auger shower depth measurements by means of new hadronic collision effects alone (if extragalactic UHECR are all protons even at the highest energies), the cross-section of proton-air interactions has to be $\sim$ 800 mb at $140$ TeV CM energy, accompanied by an increase of a factor of 2-3 in secondary particles. We also study the muon production of the showers in the same scenario. Although the muon production does increase, this increase is not enough to resolve the muon problem of UHECRs if the distribution of secondaries among different species remains unchanged with respect to the Standard Model prediction.
\end{abstract}

\maketitle


\section{Introduction}

Cosmic Rays (CR) are the most energetic particles in the universe. Over 100 years have passed since they were first discovered by Hess \cite{Hess:1912srp}, yet still today their composition, origin and acceleration mechanism are subjects of debate. This lingering uncertainty stems not only from  observational limitations but also from particle-physics uncertainties: the first collisions of ultra-high-energy cosmic rays (UHECRs $E\gtrsim 10^{18}$ eV)  with the Earth's atmosphere occur at center-of-mass (CM) energies exceeding $40$ TeV and reaching $300$ TeV; for comparison, current  lab tests of hadronic physics (in the Large Hadron Collider, LHC) only reach CM energies of 14 TeV. 

Observationally, the greatest challenges in studying UHECRs
are their low flux and their inability to penetrate the atmosphere: detection has to be indirect, tracing either the development of the extensive particle air shower (EAS) caused by a CR's collision with the atmosphere, or the EAS products reaching the ground, or both; and dedicated UHECR observatories need collective areas of thousands of km$^2$. 

Despite significant progress in recent decades in improving statistics of UHECR detections thanks to very large facilities such as the Pierre Auger Observatory in Argentina \citep{AugerInstrument} and the Telescope Array in the United States \citep{TelescopeArrayInstrument}, important questions regarding UHECR astrophysics remain open. In particular, the distribution of UHECR arrival directions on the sky, their origin, and their composition constitute three persistent, interconnected puzzles.

Because cosmic rays are charged, it is not possible to directly associate their arrival directions with the sources that accelerate them. However, there are astrophysical arguments about possible classes and cosmic locations of cosmic-ray sources, with important implications for the  resulting cosmic-ray properties. 

At lower energies ($<10^{12}$ eV), it is fairly certain that cosmic rays originate in Galactic sources. This is evidenced by differences in the cosmic ray fluxes estimated for different galaxies through observations of gamma rays originating in the decay of neutral pions produced through collisions of cosmic rays with interstellar gas (e.g., \citep{Sreekumar93,LMC, M31M33}). At very high energies ($>8\times10^{18}$ eV) we are equally confident that cosmic rays are extragalactic, since the anisotropic distribution of arrival directions that starts emerging at these energies is not correlated with the Galactic plane or the Galactic center \citep{Auger_Dipole}.

The exact energy at which the transition from Galactic to extragalactic cosmic rays occurs is still under debate (see, e.g., \citep{Gabici_Review} for a recent review), however most recent works assume that it happens somewhere between $10^{17}$ and $10^{18.5}$ eV. Hints for this are seen both in the spectrum and in the composition of cosmic rays at this energy range. The steepening of the spectrum seen at the "second knee" and the transition, at the same energies, to a heavier composition \citep{KASCADE-Grande-2019} point towards a population originating in magnetically-confining accelerators reaching its maximum possible energy (as per the Hillas criterion, \citep{HillasOriginal,KoteraOlintoReview}).

At energies between $\sim 10^{17}$ eV and $\sim 10^{18.5}$ eV, composition-sensitive observables from KASCADE-Grande, Auger, and Telescope Array indicate a transition back to lighter composition \citep{PierreAuger:2016use,KASCADE-Grande,Bergman:2020izr}. At $\sim 10^{18.7} {\rm \, eV}$ the spectrum also becomes harder, a feature known as the ankle  \citep{AugerSpectrum2010, Bergman:2020izr}.  There are two interpretations of the ankle. In the first,  the ankle marks the transition to UHECR of extragalactic origin (e.g.,~ \citep{Nagano_1992,APO2005, APO2007}). In the second, this transition is assumed to have already happened at somewhat lower energies. As a result, the composition is already light at the ankle, and the ankle spectral feature is actually a "dip" caused by electron-positron losses (e.g., \citep{BG1988,BGG2006,ABB2014}). 

The debate of ankle-versus-dip is not, however, the only controversy at these energies. A second one is the so-called composition problem of UHECR. This is summarized as follows: At energies of $~10^{18.7}$ eV, composition-sensitive variables, taken at face-value, indicate a transition back to heavier composition \citep{AugerICRC2019}. However, there are certain astrophysical indicators {\em against} a heavy composition: (a)  The spectrum at this energy is transitioning to a shallower slope, not a steeper one - i.e, there is no coincident spectral indication that the UHECR accelerators are reaching their maximum energy \citep{NewFeature2020} (b) Anisotropies start to emerge at these energies \citep{Auger_Dipole, diMatteo:2021b6, Biteau:2021eL,TA_hotspot_journal, TelescopeArrayAniso2020}. This might not be so severe a  problem if the Galactic magnetic field has the overall low strengths indicated, e.g., by \citep{JF2012}. However, recent studies of the Galactic magnetic field have shown that it is approximately an order of magnitude stronger than previously thought \citep{2019ApJ87338T} in a small region near the reported hotspot from TA \citep{TA_hotspot_journal,TelescopeArrayAniso2020}. If indeed the average Galactic magnetic field is proven to be just a few times stronger than the existing models, combined with the dipole anisotropy at high energies, we can conclude that UHECR are light nuclei. The reason is that heavy nuclei are strongly deflected from Galactic magnetic fields and would spread over all the sky, eliminating all evidence of anisotropy. (c) Heavier nuclei photodissociate fast during propagation (e.g., \citep{Puget1976,Stecker_1999, APO2007}) -- with the exception of iron -- so the composition becomes lighter during propagation, unless it starts out as pure iron. However, iron is far from a best-fit to Auger composition-sensitive observables.  Instead, observations can be better fit by a mix of intermediate-mass nuclei, requiring an astrophysically contrived composition of the accelerated particles at the source (e.g., \citep{Piran2013,AugerXmaxDist, AugerCombinedFit2017, Arsene:2021inm, AugerCombinedFitICRC2022}). In contrast, models that are more natural astrophysically are not in as good agreement with composition-sensitive observables \citep{PhysRevD.92.123001,PhysRevD.92.021302}.

In addition to these astrophysical considerations, there are particle-physics considerations that add to the composition problem.  Hybrid detectors such as Auger and Telescope Array measure composition indirectly, in two ways: from the atmospheric slant depth at which the shower reaches a maximum, $X_{\rm max}$ (measured through fluorescent detectors); and from the number of muons reaching the ground (measured by surface array tanks). At a fixed primary CR energy, heavier nuclei will typically give a lower and less variable $X_{\rm max}$; and they will produce more muons. The observations of $X_{\rm max}$ and muon numbers are then compared against the predictions from air-shower simulations. However, the best-fit compositions from muons and $X_{\rm max}$ do not match \citep{AugerMuons2019}: too many muons are produced on the ground compared to what would be expected from the best-fit composition obtained from $X_{\rm max}$ alone.

The latter two problems indicate that the air-shower simulations (or rather, the hadronic collision simulation models on which these are based) may not be capturing correctly the development of showers. This is not altogether unexpected, since the first collision of a $10^{17} {\rm \, eV}$ cosmic ray with a stationary atmospheric proton is already at  energies above those attainable at the Large Hadron Collider (LHC)\footnote{ From now on this energy regime will be called super-LHC energies}: we are simulating collisions of these primaries with the atmosphere based on theoretical extrapolations of hadronic behavior to higher energies. This has led several authors to hypothesize that the problem may lie in the hadronic collision models themselves. The solution that has been proposed in this context is that, above  a threshold energy $E_\text{th}$, the proton-Air interaction changes due to new new physics beyond the SM. This scenario is widely recognized both by the Auger Collaboration \citep{PhysRevLett.117.192001,Veberic:2017hwu,Auger2018} and other authors \citep{Ulrich:2010rg,
Farrar:2013sfa,PhysRevD.95.063005,PhysRevD.95.095035,AllenFarrar2013, 
Farrar_RSCM19,pavlidou_tomaras_2019}. In this scenario, the  composition of the primary remains light.

In  \citep{pavlidou_tomaras_2019} (hereafter PT19) we calculated analytically phenomenological constraints on any new effect that would alter hadronic interactions in such a way as to mimic a heavy composition at the highest UHECR energies  through changes in first-collision multiplicity and cross-section alone. Note that this is a different mechanism than the one evoked in  \cite{AllenFarrar2013}, who investigated the extent to which tensions between composition-sensitive observables could be alleviated by suppression of neutral pion production, either throughout the shower or in high-energy interactions only. Our work in PT19 also offered constraints only on the outcome of any mechanism altering the number of first-collision products and the proton-air cross section, in contrast with works (e.g., \cite{Farrar:2013sfa, PhysRevD.95.063005}) that discuss a specific mechanism that might result in such or different phenomenological changes in the shower development.

 In PT19 we showed that if the multiplicity of first-collision products increased over the SM predictions at a certain rate, the growth of the average $X_{\rm max}$ with energy can be fully explained while keeping the composition light even at the highest energies. We also showed that a simultaneous increase in the proton-air cross-section  over the SM prediction would improve agreement of $\sigma_{X_{\rm max}}$ (the shower-to-shower variation in $X_{\rm max}$) with the data, although we did not calculate the optimal behavior of the cross-section to best match Auger observations.  Importantly, this scenario is insensitive to the exact energy at which new Physics sets in, as long as the transition happens {\em below} the break observed in the $\langle X_{\rm max} \rangle$ vs energy plot in Auger data, and above a center-of-mass energy so low that the new effect would have been registered in LHC data. Rather, the $\langle X_{\rm max} \rangle$ break is due to the transition from Galactic to extragalactic cosmic rays, while the extragalactic cosmic ray composition remains light. 

In this paper, we extend the analytic formulation of PT19 using EAS simulations. For that purpose, we use a widely known program: CORSIKA\footnote{CORSIKA version 7.7402} \citep{1998cmcc.book.H}. CORSIKA uses  extrapolations of SM at  post-LHC energies to model hadronic interactions. In this work we used two such models: EPOS LHC \citep{Pierog_2015} and QGSJETII-04 \citep{PhysRevD.83.014018}. For low energy interactions we used FLUKA \footnote{FLUKA version 2020.0.3} \citep{Ferrari:2005zk} which is a Monte Carlo code used extensively at CERN. Our main goals are: (a) To test whether an increase in the multiplicity of first-collision products can indeed yield the changes in the $X_{\rm max}$ distribution predicted in PT19 -- as contrasted with, e.g. the implementation of such a change by \citep{Ulrich:2010rg}, who found that the variance of $X_{\rm max}$ remains practically unchanged under a change in product multiplicity.  In contrast, in our work it is the increase in product multiplicity that drives {\em all} of the deviation of $\langle X_{\rm max}\rangle$ from the Standard Model predictions, and is also responsible for a large fraction of the effect on $\sigma_{X_{\rm max}}$ (in contrast to other implementations of new physics at the highest-energy UHECR collisions, e.g. \cite{Farrar_RSCM19}, where the impact on composition-sensitive observables is driven by cross-section effects). (b) To calculate the optimal change in cross-section that best matches Auger $X_{\rm max}$ data. In PT19 we did not fully explore the parameter space, but we only argued that an increase in cross section changes the  $X_{\rm max}$ variance in the direction of better agreement with Auger data.   The behavior of the multiplicity on the other hand was set in PT19  by the requirement to fit the high-energy branch of the scaling of $\langle X_{\rm max} \rangle$ with energy; and this is also true in the present work. (c) To evaluate the impact of this scenario on the muon problem, which was not addressed in PT19. {(d) To expand on the Galactic-to-extragalactic transition, which, in the scenario we consider, is the source of the break observed in Auger data in the $\langle X_{\rm max} \rangle$ versus energy plot. }

This paper is organized as follows.   In Section \ref{form} we present the implementation of new physics effects in the first collision of $E>10^{18} {\rm \, eV}$ CR with the atmosphere.  In particular, we discuss the way we parameterize changes in the multiplicity and cross section in In \S \ref{sec1}. We show how these changes impact the slant depth of the showers in  \S \ref{sec1.1}, and the muons reaching the ground in \S \ref{sec2}. In \S \ref{sec4} we describe how we implement these changes of \S \ref{sec1} in air shower simulations using CORSIKA.
In this paper, we assume that all extragalactic cosmic rays reaching the Earth are protons. However, not all $E>10^{18} {\rm eV}$ cosmic rays detected on Earth  are extragalactic, and, more importantly, the high-energy-end of the Galactic cosmic ray spectrum has a heavy composition, additionally affecting the slant depth and muon content. In order therefore to compare our results with observations, we additionally need a model for the way the Galactic cosmic ray flux cuts off with energy. We describe this model in \S \ref{sec3}. 
We discuss the results of our simulations in Section \ref{sec5}, and we summarize and discuss our conclusions in Section  \ref{sec6}. 

\section{Implementation of the "new Physics above 50 TeV" scenario}\label{form}

The "new physics above 50 TeV" scenario that we explore here using simulations of EAS implements two phenomenological changes in the {\em first collision} of the incoming primary cosmic ray with the atmosphere (as in PT19):
(1) an increase in the multiplicity of the first collision products; and 
(2) an increase in the cross section of the first interaction with the atmosphere. Our approach is phenomenological and not tied to any specific new physics model. However,  several candidate particles and new physics mechanisms exist that might lead to such  behavior (see, for example, \citep{ms2005,cct2005,mmt2009,Farrar:2013sfa}).

\subsection{Parametrization of changes in cross section and multiplicity}\label{sec1}

The cross-section of protons with nitrogen has a logarithmic behaviour at high energies \citep{ReviewParticle2016}. For that reason,  the cross section is usually parameterized as
\begin{equation}\label{cross_section_unchanged}
    \sigma_\text{p-Air} = \sigma_0 + \beta \,\log \varepsilon
\end{equation}
where $\sigma_0$ and $\beta$ are constants. Here we normalize the energy scale as $\varepsilon =  E /E_\text{th}$, where $E_\text{th}$  is the threshold energy above which new physics sets in. Based on the arguments in PT19, we will take $E_\text{th} = 10^{18}$ eV, corresponding, for a collision of an primary proton with a stationary atmospheric proton, to a CM collision energy of $\sim 50$ TeV. We note that this threshold is neither coincident with the location of the break seen in Auger $X_{\rm max}$ data, nor fine-tuned (other choices that satisfy both that the threshold is ultra-LHC and that it lies below the Auger data break give similarly good results). We can calculate $\sigma_0$ and $\beta$ from hadronic interaction models (here we use EPOS LHC and QGSJETII-04) and the standard-model extrapolations employed therein. The results are given in the first 2 lines of Table \ref{tab1}.

\begin{table}
\centering
    \begin{tabular}{|l|c|c|}
    \hline
        & EPOS LHC & QGSJETII-04\\
        \hline\hline
        $\sigma_0$ (mb)         & 527.61 & 499.35 \\
        \hline
        $\beta$ (mb)            &  49.95 & 37.49 \\
        \hline
        $X_0$ (gr$/$cm$^2$)     & 706.35 & 691.46 \\
        \hline
        $\alpha$ (gr$/$cm$^2$)  &  62.78 & 61.13\\
     \hline
    \end{tabular}
    \caption{Parameters for cross section and shower maximum. These parameters were calculated by performing linear fit in the SM data from EAS simulation with proton as a projectile.}
    \label{tab1}
\end{table}

If new phenomena take place at $\varepsilon >1$,
the cross section of the first interaction may change. Here we assume that such a change will only affect the value of the coefficient $\beta$ of the energy-dependent term, and that the cross section will be continuous at $\varepsilon = 1$:
\begin{equation}\label{cross_section}
    \sigma_\text{p-Air,new} = \sigma_0 + \beta' \,\log \varepsilon
    \end{equation}
We will parameterize this change in terms of the fractional change $\delta$ in the coefficient $\beta$ relative to its standard-model value. Defining then 
\begin{equation}\label{delta}
\delta = (\beta'-\beta)/\beta\,,
\end{equation}
we obtain 
\begin{equation}\label{eq:sigma_new}
 \sigma_\text{p-Air,new} =\sigma_\text{p-Air}+ \delta\,\beta \,\log \varepsilon\,.
\end{equation}
In other words, for $\varepsilon >1$, the cross section will deviate logarithmically from its standard-model--predicted value at the particular energy, with a coefficient $\delta \beta$. Clearly, $\delta=0$ corresponds to no change to the cross-section at any energy over the standard-model prediction. We note however that the uncertainty in the standard-model predictions for the proton-air cross-section to super-LHC energies is high (see, e.g., Fig. 2 of \citep{Ulrich:2010rg}), and values of $\delta$ as high as 3.5 could still be consistent with the standard model within uncertainties.

We also postulate an increase of the number of secondary particles produced after the first collision of the primary with the atmosphere. We limit the effect to the first collision since, for energies of interest, secondary particles will, with very high probability, have energies such that their collisions with air will occur at CM energies below the threshold for new physics. We parameterize this increase in first-collision product multiplicity by 
\begin{equation}\label{multiplicity_def}
    n(\varepsilon) = N(\varepsilon)/N_\text{SM}(\varepsilon)\,,
\end{equation}
where $N$ is the number of secondaries produced under new physics, and  $N_\text{SM}$ is the standard-model prediction for the number of secondaries. It is possible that new physics may also change the charged-particle ratio of the products; we do not however implement such a change here. We further discuss this issue in \S \ref{sec5}.

\subsection{Effect of changes in cross section and multiplicity on shower maximum}\label{sec1.1}
The slant depth of the shower maximum
is the air column density traversed by the EAS front -- measured from the top of the atmosphere -- for which the shower front achieves its maximum atmospheric ionization rate:
\begin{equation}\label{defXmax}
    X_\text{max} = \int\limits_\infty^{x_\text{max}} \rho(l)\, dl\,.
\end{equation}
In Eq.~(\ref{defXmax}), $\rho$ is the density of the atmosphere, $l$ is a length measured along the path the shower particles traverse in it, and $x_\text{max}$ is the height of the shower maximum. $X_\text{max}$ quantifies the total atmospheric column the shower has already encountered at its maximum, independently of the inclination of the incoming CR. 
$X_\text{max}$ can be written as the sum of two terms. The first one is the column density after which the first interaction of the cosmic ray primary takes place, $X_\text{int}$; and the second one is the column density between first interaction and shower maximum, $X_\text{long}$,
corresponding to the "longitudinal" development of the shower:
\begin{equation}\label{xmaxsum}
    X_\text{max}=X_\text{int} + X_\text{long}\,.
\end{equation}

\subsubsection{The first interaction: $X_{\rm int}$}

The probability that a CR has not interacted with the atmosphere in the vicinity of height $x$ is 
$$\exp\Bigg(-\frac{\sigma_\text{CR-Air}}{m}\int\limits_\infty^x\rho(l)\;dl\;\Bigg),$$
where $m$ is the average mass of the particles in the atmosphere (mainly nitrogen) and $\sigma_{\text{CR-Air}}$ is the cosmic-ray--air cross section for the energy of the primary. 
The average value of the depth of the first interaction for a given primary energy thus is 
\begin{equation}
\langle X_\text{int}\rangle = \frac{m}{\sigma_{\text{CR-Air} }(\varepsilon)}\,.
\end{equation}
Since $X_\text{int}$ follows Poisson statistics, its variance will be
\begin{equation}
    \text{Var}(X_\text{int}) = \langle X_\text{int}\rangle^2 = \frac{m^2}{\sigma^2_{\text{CR-Air}}(\varepsilon)}\,.
\end{equation}
If new physics sets in for $\varepsilon >1$, the distribution of $X_\text{int}$ for primaries of a given energy will be affected through the change in $\sigma_\text{CR-Air}$, which (assuming the composition remains light) will take the value of $\sigma_\text{p,Air,new}$ given by Eq.~(\ref{eq:sigma_new}):
\begin{equation}\label{avgintnew}
\langle X_\text{int,new}\rangle 
= \frac{m}{\sigma_\text{p-Air}(\varepsilon)+ \delta\,\beta \,\log \varepsilon}
\,,
\end{equation}
and 
\begin{equation}\label{varintnew}
    \text{Var}(X_\text{int,new}) =  \frac{m^2}{[\sigma_\text{p-Air}(\varepsilon)+ \delta\,\beta \,\log \varepsilon]^2}\,.
\end{equation}

\subsubsection{The longitudinal development: $X_{\rm long}$}
Following the CR interaction with the atmosphere, secondary and subsequent generation of particles are produced. As the shower of particles evolve though the atmosphere, the energy per particle decreases, partly because of ionization losses and partly because of new particle production.  This process continues until the energy of the shower particles drops below the energy threshold of new particle production, at which point the shower continues to evolve by ionization losses alone. This evolution of the shower is observable through the fluorescence of ionized atoms of the atmosphere, which can be detected with ground telescopes. The intensity of this fluorescent light encodes the energy loss rate of the shower front. $X_{\rm long}$ measures the column density traversed by the shower front {\em after the first interaction} until the energy loss rate reaches its maximum value. 

The dependence of the longitudinal column, $X_\text{long}$, with energy can be derived from the simple model of Heitler \citep{Heitler:1936jqw,MATTHEWS2005387}, where the initial particle produces two daughter particles which split the primary's energy, and the process continues until the energy of each particle reaches a critical energy below which the process cannot continue. 

This results to a logarithmic increase of $X_\text{long}$ with energy. A more realistic calculation of $X_\text{long}$ is more complex because additional phenomena take place (e.g. bremsstrahlung, pair production, pion production, hadronization), and there are significant shower-to-shower fluctuations. For that purpose numerical simulations are used to follow the development of EAS (e.g. CORSIKA). The final behavior of $\langle X_{long} \rangle$ with energy, however, is still logarithmic:
\begin{equation}\label{avglong}
    \langle X_\text{long}\rangle = X_0 + \alpha \, \log \varepsilon
\end{equation}
where $X_0$ and $\alpha$ are constants, while $\sigma_{X_\text{long}}$ remains approximately constant with energy. 

In the last 2 lines of Table (\ref{tab1}) we show the best-fit parameters $X_0$ and $\alpha$ for protons, derived from CORSIKA EAS simulations using two different models (EPOS LHC and QGSJETII-4) of hadronic interactions. 
Both models employ standard-model extrapolations for collisions at super-LHC energies. The parameters do depend slightly on the hadronic interactions model, however both models produce qualitatively similar results. 

If now new physics sets in for $\varepsilon >1$,  the distribution of $X_\text{long}$ will change. To quantify this change, we model empirically the shower as $n(\epsilon)$ "component showers", of energy $\epsilon/n(\epsilon)$ on average, developing independently. Note that this approach is conceptually and qualitatively different from that of \cite{Ulrich:2010rg} who used a multiplicative factor 
to increase the number of products in each collision  - the most prominent difference being that the presence of independently developing "component showers" decreases the shower-to-shower fluctuations of $X_\text{long}$ (i.e. $\sigma_{X_\text{long}}$), while a multiplicative increase of products of identical distribution as the original shower leaves $\sigma_{X_\text{long}}$ unchanged.

Under this change, $\langle X_{\rm long, new} \rangle$ becomes  [from Eq.~(\ref{avglong})]
\begin{equation}\label{avglong_new}
    \langle X_\text{long,new}\rangle = X_0 + \alpha \, \log \frac{\varepsilon}{n(\epsilon)}\,.
\end{equation}
To produce an analytical estimate the variance of $X_\text{long}$ we take, as in PT19,  the average of individual "component-shower" longitudinal depth,  $\cfrac{1}{n}\sum\limits_i X_{\rm long,i}$ to be a reasonable estimation of $X_{\rm long}$. Then $X_{\rm long}$ is the “sample mean” of $n$ “draws” from the underlying distribution of $X_{\rm long,i}$ and the distribution of these “sample means” has a variance that is given
by the “error in the mean” formula, 
\begin{equation}\label{varxlongnew}
    \text{Var}(X_\text{long,new}) = \frac{\text{Var}(X_\text{long,i})}{n(\varepsilon)}\,.
\end{equation}
Here $\text{Var}(X_\text{long,i})$ is the variance of $X_\text{long,i}$, and it can be assumed to follow the SM predictions, since each subshower of the sample will have $\epsilon <1$. 

\subsubsection{New physics mimics a transition to heavy composition}
The average value and the variance of $X_\text{max}$ are sensitive to the first interaction - both its cross-section and the multiplicity of its products. A higher first-interaction cross-section will result to lower $\langle X_\text{int} \rangle$ and  $\text{Var}(X_\text{int})$, and therefore lower $\langle X_\text{max} \rangle$. A large number of first collision products will distribute the energy of the primary more widely, resulting to lower $\langle X_\text{long} \rangle$ and therefore to lower $\langle X_\text{max} \rangle$. A larger number of first-collision products will also result to reduced shower-to-shower fluctuations in $X_\text{long}$ and in turn to a lower $\text{Var}( X_\text{long})$ and a lower $\text{Var}( X_\text{max})$. 

A heavy primary composition will drive both $\langle X_\text{max} \rangle$ and $\text{Var}(X_\text{max} )$ to lower values for a given energy, through both an increased cross-section of the primary-air collision, and an increased first-collision product multiplicity. The new-physics scenario we discuss here will also move the $X_\text{max}$ distribution in the same direction through similar changes in cross-section and multiplicity, and can thus mimic a transition to a heavier composition. 

\subsubsection{Constraining $n(\varepsilon)$ and $\delta$}
The empirical model we have presented here features a {\em free parameter} $\delta$ and a {\em free function} $n(\varepsilon)$. However, if we assume that the change of slope in $\langle X_\text{max} (\epsilon)\rangle$ observed by Auger at the highest energies is, {\em in its entirety}, due to new physics, so that all primaries at these energies are protons, we can completely determine $n(\epsilon)$ as a function of $\delta$, leaving only a single free parameter in our model (which can also be optimized by comparison to other moments of the $X_\text{max}$ distribution, as we will see in \S \ref{sec5}). 

The Auger Collaboration reports, for energies $E\geq 10^{18.3}$ eV
\begin{equation}
    \langle X_\text{max, Auger} \rangle = X_{0,\text{ Auger}} + \alpha_\text{Auger} \, \log \varepsilon 
\end{equation}
with $X_{0,\text{Auger}} = 742.55\;\text{gr}/\text{cm}^2$ and $\alpha_\text{Auger} = 23.98\;\text{gr}/\text{cm}^2$. Simulated data of pure protons air showers indicate a difference from Auger observations above this energy.
By equating the behavior of Auger data to the results of the postulated new physics effects at $\epsilon > 1$ [using Eqs.~(\ref{xmaxsum}), (\ref{avgintnew}), and (\ref{avglong_new})] we obtain 
\begin{align}
    \langle X_\text{max,Auger}\rangle (\varepsilon) = \cfrac{m}{\sigma_\text{p-Air} (\varepsilon)+ \delta\,\beta \,\log \varepsilon} \nonumber\\
    + X_0 + \alpha \, \log \cfrac{\varepsilon}{n(\varepsilon)}
\end{align}
and solving for the multiplicity we obtain
\begin{align}\label{multiplicity}
    \log n(\varepsilon) = \frac{X_{0} -X_{0,\text{Auger}}}{\alpha} &+ \cfrac{\alpha - \alpha_\text{Auger}}{\alpha} \log \varepsilon \nonumber\\
    &+ \cfrac{1}{\alpha}\cfrac{m}{\sigma_\text{p-Air}(\varepsilon)+ \delta\,\beta \,\log \varepsilon}
\end{align}
with $\sigma_\text{p-Air}(\varepsilon)$ given by Eq.(\ref{cross_section_unchanged}). 

The variance of $X_{max}$ under new physics effects can be estimated through
\begin{equation}
    \text{Var}(X_{max,new}) =  \text{Var}(X_\text{int, new})+  \text{Var}(X_\text{long,new})
\end{equation}
using Eqs.~(\ref{varintnew}) and (\ref{varxlongnew}), yielding
\begin{align}\label{varxmaxnew}
   \text{Var}&(X_{max,new}) = \sigma_{X_\text{max}}^2 \nonumber\\
   &= 
   \frac{m^2}{[\sigma_\text{p-Air}(\varepsilon)+ \delta\,\beta \,\log \varepsilon]^2}
+ \frac{\text{Var}(X_{long})}{n(\varepsilon)}\,, 
\end{align}
which again has only one free parameter: $\delta$. Clearly, $\delta$ can be optimized by comparing Eq.~(\ref{varxmaxnew}) with Auger data on $\sigma_{X_\text{max}}$.

\subsection{Effects of changes in cross section and multiplicity on the number of muons on the ground}\label{sec2}
The muonic part of EAS is also a problem in UHECR physics. Muons are produced mainly when charged pions or kaons decay, indicating hadronic interactions. They have a large mean free path and consequently, their journey in the atmosphere is mostly undisturbed. Upon reaching the ground, muons can be detected via the Cherenkov radiation they produce inside the water tanks comprising the surface array of hybrid experiments such as Auger and TA. Thus the energy and spatial distribution of muons can be measured.

A useful parameter used to compare experiments with simulations is the ratio
\begin{equation}
    R_\mu=\cfrac{N_\mu}{N_{\mu,19}}
\end{equation}
where $N_\mu$ is the number of muons detected on the ground. The reference parameter $N_{\mu,19}=1.455\cdot 10^7$ is inferred from simulations assuming proton as the primary particle with energy of $10^{19}$ eV, taking into consideration the detector's response for muons above $0.3$ GeV \footnote{This is the Cherenkov threshold for Auger's water tanks.} that reach Auger's site at an altitude of $1425$ m with an inclination of  $\theta = 67^\circ$. \citet{PierreAuger:2021qsd} reports that $N_{\mu,19}$ does not depend strongly on the selected high energy model. They quote a $\sim 11\%$ systematic error introduced in this way. The reason for selecting so inclined showers is because for inclination $\theta>60$, EAS are dominated mostly by muons, since the electromagnetic part is absorbed by the atmosphere.  We used $67$ degrees as zenith angle for our shower simulations.

Auger reports that the average of that ratio depends on energy as
\begin{equation}
    \langle R_\mu\rangle= a\Bigg(\cfrac{E}{10^{19}\text{ eV}}\Bigg)^b
\end{equation}
where $a = 1.86$ and $b=0.99$. 

 Auger observes $25-40\%$ more muons at $10^{19}$ eV than high energy models predict, assuming proton as a primary. However, muons are observed to be in overabundance {\em even if a heavier composition, consistent with the one that would produce the observed $\langle X_\text{max}\rangle (\epsilon)$ is assumed for the primaries}.

In the scenario where proton interactions change above $10^{18}$ eV introducing new physics, the muon production will change due to the increase of multiplicity for the secondary particles. 

The total number of muons on the ground will be the sum of muons that each component-shower produces. Note that each component-shower does not produce the same amount of muons. 

We then expect the average number of muons on the ground to be
\begin{equation}
    \langle N_{\mu,\;new}(\varepsilon)\rangle = n(\varepsilon)\;\langle N_{\mu}(\varepsilon/n)\rangle\,.
\end{equation}

\subsection{Implementation of altered multiplicity and cross section in CORSIKA simulations}\label{sec4}

We now  turn to EAS simulations. We wish to: (a) test whether
the implementation, in simulations, of new physics as discussed in \S \ref{sec1} will produce the same behavior as our analytic approximations; and (b) determine the optimal phenomenological parametrization (cross section and the multiplicity, quantified by $\delta$ and $n(\varepsilon)$ in our description) that any new proton-air interaction must exhibit in order for EAS to produce the observed data in Auger if the composition of primaries is to remain light up to the highest energies. 

We simulated showers induced by primaries with energies in the range $10^{17}-10^{20}$ eV with step in $\log E$ of $0.1$. At each energy  bin, we performed $1000$ EAS simulations. We simulated EAS with the first collision treated either with SM extrapolations or with new physics as per our phenomenological implementation. We performed SM EAS simulations for proton primaries with $E<E_{\rm th} = 10^{18} {\rm \, eV}$, and for the heavier Galactic primaries (see next section), since their per-nucleon kinetic energy never exceeds $10^{18} {\rm \, eV}$. We implemented new Physics for all proton primaries with $E>10^{18} {\rm \, eV}$  in the first collision only, i.e. we assume that all first-collision products have center-of-mass energies below the new-Physics threshold in all subsequent collisions.

CORSIKA EAS simulations using either EPOS-LHC or QGSJETII-04 yielded our SM results. For low-energy interactions we used FLUKA. We also used the CONEX hybrid scheme \citep{Bergmann:2006yz} which decreases the simulation time dramatically. Each simulation generates three output files. The first records the energy deposited as a function of depth in the atmosphere. We fitted a Geisser-Hillas function
\begin{align}\label{GH-function}
   \cfrac{dE}{dX}(X) = N_{max}\Bigg(\cfrac{X-W_0}{X_\text{max}-W_0}\Bigg)^{\frac{X_\text{max}-W_0}{\lambda}} \nonumber\\
   \times\exp\Bigg(\cfrac{X_\text{max}-X}{\lambda}\Bigg)
\end{align}
to the simulated data to estimate the shower longitudinal-development maximum, $X_\text{long}$, for each shower.  $W_0$ and $\lambda$ are two fitting parameters.  In our simulations we fixed the height of the interactions at 60km. Thus in Gaisser-Hillas function $X_\text{max}$ becomes $X_\text{long}$, the distance from the first interaction. This was done in order to have control over the $X_\text{int}$.  For each energy bin, we then calculated the average value of  $ X_\text{long}$ and its variance. 

The second output file records information on the cross section of the primary with the atmosphere and from it we calculated $\langle X_\text{int}\rangle$. The second output file  also contains the number of muons detected on the ground. We fitted a convolution of a Gaussian with an exponential to calculate the average value and the relevant variance of the number of muons on the ground. The third output file (stack file) contains information about the secondary particles produced after the first interaction. The results from this run are the Standard Model predictions from extrapolated models. 

For the new-physics simulations, we used the following approach. For a given value of $\delta$, we first  calculated the multiplicity according to Eq.(\ref{multiplicity}) at each energy bin. We then combined stack files from the same energy bin (produced by SM simulations) according to the calculated multiplicity: we rounded $n(\varepsilon)$ to the nearest integer, and we combined as many stack files, accounting for energy and momentum conservation. To  this end, we divided the energy and momentum of each particle by the number of stacked files. We then used the combined stack files as input to CORSIKA and continued the simulation of the EAS, obtaining data files for the energy deposition as a function of depth in the atmosphere and the muon number on the ground. The cross section was calculated from Eq.(\ref{cross_section}). 

\subsection{Transition from Galactic to extragalactic \\ cosmic rays and imprint on $X_{\rm max}$}\label{sec3}

\begin{figure}[htbp]\vspace{-10pt}
   \centering
   \includegraphics[width=0.53\textwidth]{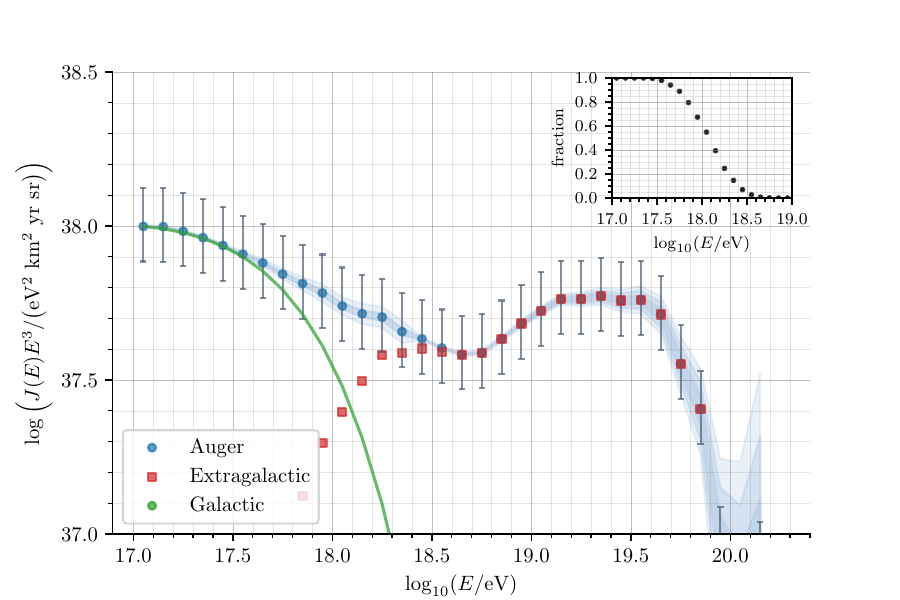}
      \caption{CR energy spectrum (flattened by $E^3$). Blue:  Auger Collaboration data. Green: Galactic CR flux model, assuming  power law at low energies with an exponential cutoff at $10^{17.9}$ eV. Red: extragalactic CR flux, obtained as the difference between observed data and Galactic CR. Inset: fraction of Galactic to total CR flux}
         \label{fig:fig1}
\end{figure}

At energies above $10^{18.3}$ eV, both Auger and TA observe a dipole distribution of CR uncorrelated with the galactic plane \citep{Aab:2018mmi,Abbasi:2020ohd}. This indicates that above this energy, CR are of extra-galactic origin. The energy at which  this transition takes place is an important input for our calculations.   There are two reasons for this: First,  Galactic CR are heavy particles and thus the energy per nucleon will be below the threshold of new physics. As a result we need not apply any new physics corrections to their EAS simulations.  Second, in the scenario we are considering here, the break in $\langle X_{\rm max} \rangle$ is a result of the Galactic-to-extragalactic transition rather than the transition to new Physics. The new effect enters at a considerably lower energy; that energy is not registered in the spectrum due to Galactic cosmic rays dominating and the proton abundance in them being extremely low.

We model this transition using a simple, phenomenological approach, based on three assumptions: (1) That above $10^{17}$ eV cosmic rays consist of a single, fixed-composition Galactic component, and a single, fixed-composition extragalactic component. The energy-dependent fraction of Galactic CR is $f(\varepsilon)$. The fraction of extragalactic CR is then $1-f(\varepsilon)$. (2) That above $10^{17}$ eV we can model the Galactic CR spectrum (differential particle flux $J(\varepsilon)$) as a power law of slope $-\gamma_G$, cutting off exponentially at a a characteristic energy $\varepsilon_G$, corresponding to the maximum energy of Galactic CR accelerators: 
\begin{equation}
J_G(\varepsilon)=J_{G,0}\Bigg(\cfrac{\varepsilon}{\varepsilon_{17.5}}\Bigg)^{-\gamma_G}\exp{\Big(-\varepsilon/\varepsilon_G\Big)}
\end{equation}
where $\varepsilon_{17.5}= 10^{17.5}\text{eV}/E_\text{th} = 10^{-0.5}$. 
(3) That CR at energies lower than those where losses (either e$^+$e$^-$ or pion photoproduction) become important, the extragalactic CR flux $J_{EG}(\varepsilon)$
 is a single power law. 
 
 At low energies ($E<10^{17.5}$ eV), Auger data constrain  
$J_{G,0}=4.1\times 10^{-15}\,(\text{km}^2\;\text{eV}\;\text{yr}\;\text{sr})^{-1}$, and  $\gamma_G=2.9$. By virtue of our third assumption above, we can also constrain $\varepsilon_G$ by demanding that, for $10^{17.5} \text{eV}/E_\text{th}<\varepsilon< 10^{18.2} \text{eV}/E_\text{th}$ the extragalactic spectrum $J_{EG}(\varepsilon) = J_\text{total, Auger} - J_G(\varepsilon)$ is consistent with a single power law. We thus find $\varepsilon_G = 10^{17.9} \text{eV}/E_\text{th}$, which results in an extragalactic spectrum consistent with $J_{EG} (\varepsilon) \propto \varepsilon^{-2.0}$ between $10^{17.5}$ and $10^{18.2}$ eV (see Fig.~\ref{fig:fig1}). 

The resulting Galactic CR fraction, $f(\varepsilon) = J_G(\varepsilon) / J_\text{total,Auger}(\varepsilon)$ is shown in the inset of Fig.~\ref{fig:fig1}. We note that under the three assumptions adopted here, extragalactic CR are found to dominate already at $10^{18}$ eV, the composition at $10^{18.5}$
 is light, and the ankle must be an e$^+$e$^-$ "dip". In this simple scenario, the probability density function of $X_\text{max}$ will be
\begin{equation}
    p(X_\text{max}) = f\;p_G(X_\text{max}) +(1-f)p_{EG}(X_\text{max})
\end{equation}
leading to an average shower maximum
\begin{equation}
    \langle X_\text{max}\rangle = f\;\langle X_\text{max}\rangle_G +(1-f)\langle X_\text{max}\rangle_{EG}
\end{equation}
and its variance
\begin{align}
    Var(X_\text{max}) &= f\;Var(X_\text{max,G}) \nonumber\\
                    &+(1-f)\;Var(X_\text{max,EG})\nonumber\\
            &+f(1-f)\;(\langle X_\text{max}\rangle_G-\langle X_\text{max}\rangle_{EG})^2
\end{align}
with subscripts G and EG referring to the Galactic and extragalactic populations respectively.

At energies around $10^{17}$ eV, CR are mainly of Galactic origin. In this paper we assume Galactic CR to be one type of nucleus for simplicity. We assume the Galactic component to be helium for simplicity -- lighter than PT19, who had assumed carbon. Ultimately,  the Galactic component should ideally be simulated with appropriate mixed composition, with each species cutting off at different energies according to its charged, as per the Hillas criterion \citep{HillasOriginal}. 

\section{Results}\label{sec5}

\begin{figure}%
    \centering
    [\centering QGSJETII-04] {{\includegraphics[width=0.5\textwidth]{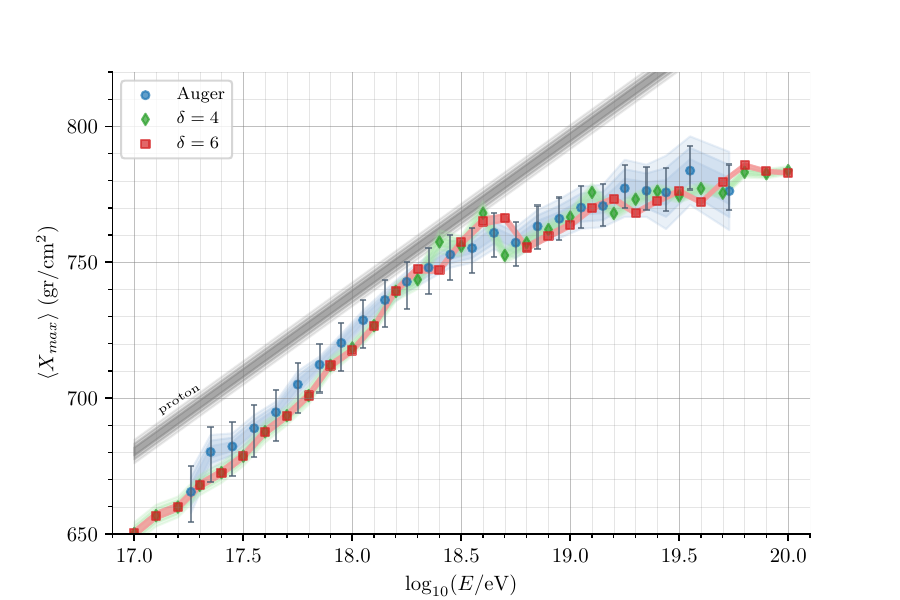} }}%
    \hspace{-10pt}
    [\centering EPOS LHC] {{\includegraphics[width=0.5\textwidth]{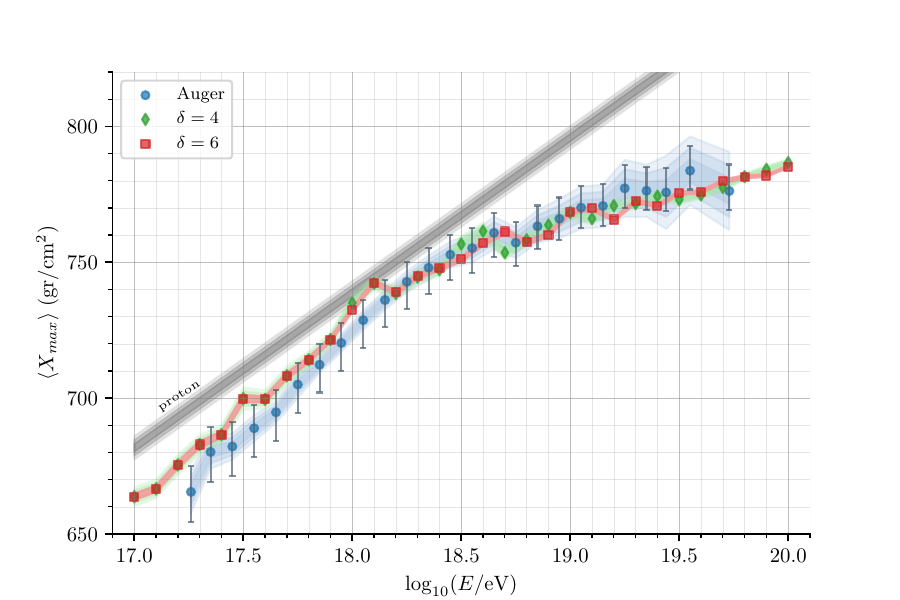} }}%
    \caption{Average shower maximum as a function of energy. Standard model extrapolations (grey) through QGSJETII-04 (upper) and EPOS LHC (lower) do not agree with observations from Auger Observatory (blue) above $E_\text{th}=10^{18.5}$ GeV. When we alter the way the cross section and the first-interaction product multiplicity scale with energy as in Eqs.~(\ref{eq:sigma_new}) and (\ref{multiplicity}), EAS simulations with protons as a primaries (green and red) reproduce the observed data well at the highest energies.The simplifying assumption of a single component Galactic CR is the reason our simulations deviate from observational data at lower energies.  In the upper plot there is a feature at $10^{18.6}$ eV. This is caused by the rounding of the multiplicity number in the creation of the first-collision secondaries files.}
    \label{fig:3}%
\end{figure}

\begin{figure}[htbp]%
    \centering
    [\centering QGSJETII-04] {{\includegraphics[width=0.5\textwidth]{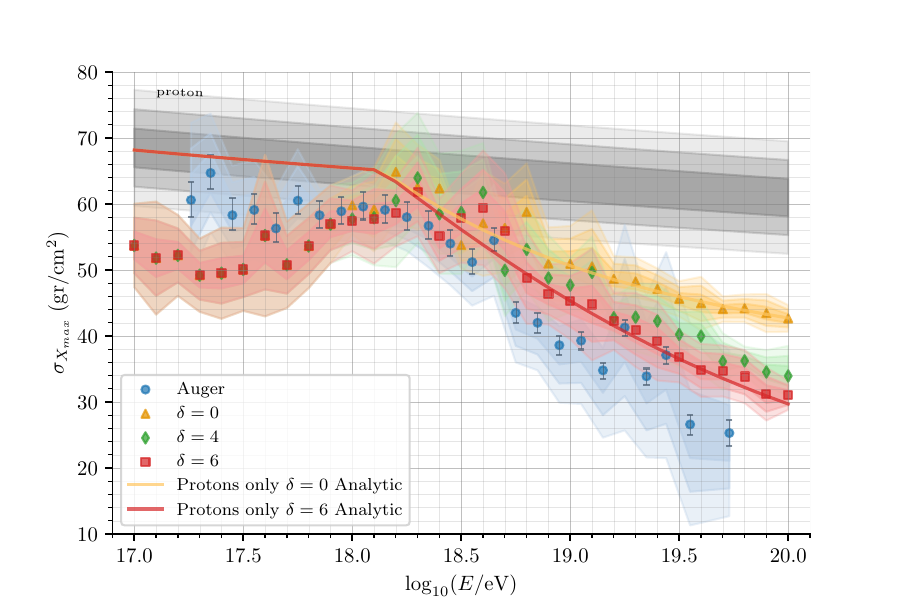} }}%
    \hspace{-10pt}
    [\centering EPOS LHC] {{\includegraphics[width=0.5\textwidth]{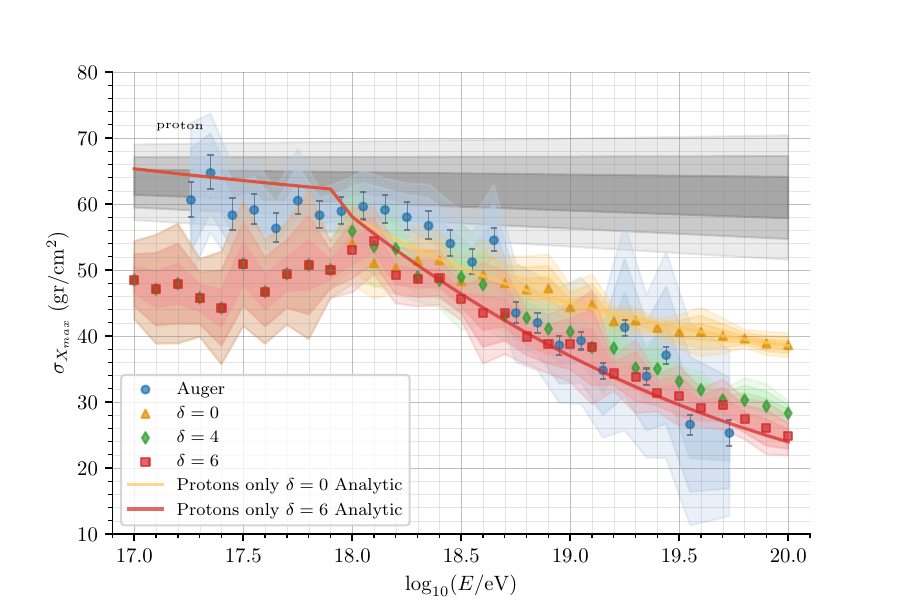} }}%
    \caption{Standard deviation of shower maxima as a function of energy. Although the mean of $ X_\text{max}$ does not depend on the parameter $\delta$ (see Fig.~\ref{fig:3}), its standard deviation does. We obtain the best agreement between simulated $\sigma_{X_\text{max}}$ and Auger  observations for $\delta$ near $6$ for QGSJETII-04 (upper) and for $\delta$ near $4$ for EPOS-LHC (lower). Further increment of the $\delta$ parameter does not result in significant changes at high energies, as $\sigma_{X_\text{max}}$ quickly reaches an asymptotic behavior. }%
    \label{fig:4}%
\end{figure}

We performed CORSIKA simulations as described in \S \ref{sec4} for $\delta=0, 2.9, 3.5, 4, 6, 8$ and $10$, treating the Galactic-to-extragalactic transition as described in \S \ref{sec3}.  These values are selected as follows: 
 $\delta=0$ is our control where there is no change in cross-section (but there is still a change in the final result due to the increase in multiplicity);  $\delta=2.9$ was the best-fit value determined analytically in PT19; $\delta=4$ is our best fit;  $\delta=3.5$ was added as an intermediate value between the PT19 best-fit and the current best-fit, to better understand the behaviour close to the optimal value;  $\delta=6$ was selected to observe the asymptotic behaviour for large $\delta$. The simulation results for $\langle X_\text{max} \rangle$
 are (by construction) in excellent agreement with Auger data\footnote{Throughout this paper, we use error bars for systematic uncertainties, and shaded areas to indicate 1, 2, and 3$\sigma$ statistical uncertainties.} for all values of $\delta$ (Fig.~\ref{fig:3}). Our simulations are also a much improved fit to the $\sigma_{x_\text{max}}$ data even for $\delta=0$ (no change in cross section, only multiplicity increases). For higher $\delta$ the agreement improves further and is optimal for $\delta$ between 4 and 8 (Fig.~\ref{fig:4}). In  Figs.~\ref{fig:3} and \ref{fig:4} we did not plot results for all simulated values of $\delta$ in order to keep the figures legible.  

Figs.~\ref{fig:3} and \ref{fig:4} show a deviation of the simulation results from observations at low energies. We expect that this discrepancy is due to the assumption of a Galactic component consisting purely of helium. A more reasonable assumption is a mixture of helium, carbon and oxygen, with ratios that depend on the energy. However, since the details of Galactic CR composition have little to no impact on the (dis)agreement between theory and observations of $X_\text{max}$ at the highest energies, we chose, for simplicity, not to focus on the modeling of the Galactic component in this work. Instead, we implement the very simple recipe described above aiming only to show the direction in which $\langle X_{\rm max}\rangle$ and $\sigma_{X_{\rm max}}$ will change due to the Galactic-to-extragalactic transition. We plan to return to this problem and relax the assumption of a single-species Galactic component in a future publication. 

\begin{figure}%
    \centering
    \includegraphics[width=0.5\textwidth]{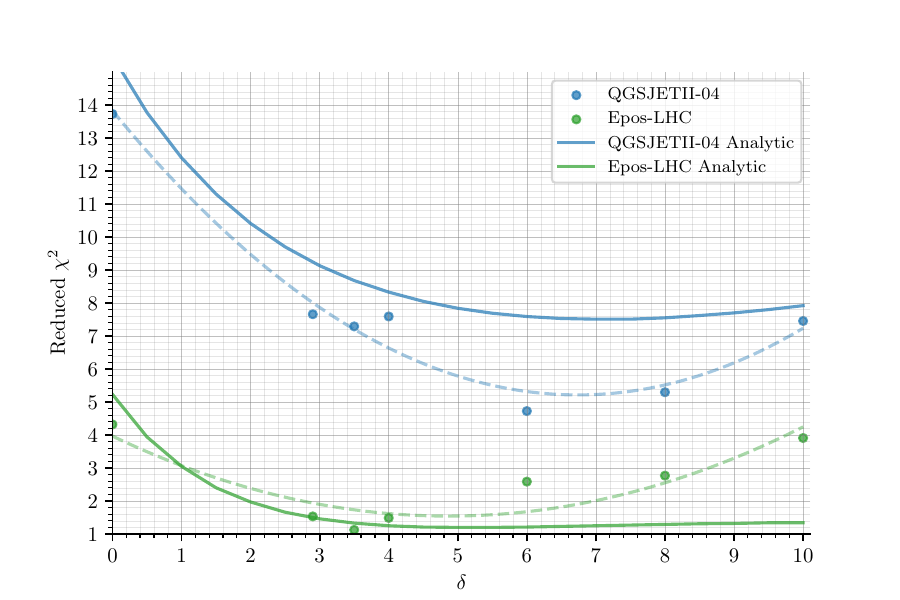}
    \caption{Agreement between new-physics EAS simulations and Auger data for $\sigma_{X_\text{max}}$ and $E>10^{18.5} {\rm \, eV}$, quantified through the reduced $\chi^2$ statistic, as a function of the value of the $\delta$ parameter. Overall, EPOS-LHC simulations produce results more consistent with observational data. To find the position of the minimum $\chi^2$, we perform a parabolic fit to the datapoints. The locations of the minima are at $\delta=4.5$ for EPOS-LHC and at $\delta=6.7$ for QGSJETII-04.}%
    \label{fig:2}%
\end{figure}

\begin{figure}%
    \centering
    {{\includegraphics[width=0.5\textwidth]{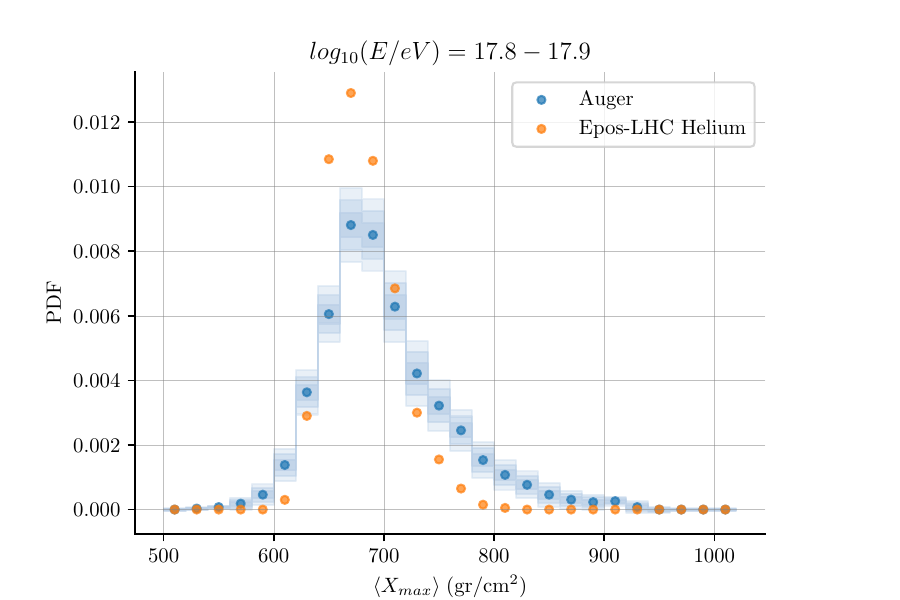} }  }%
    {{\includegraphics[width=0.5\textwidth]{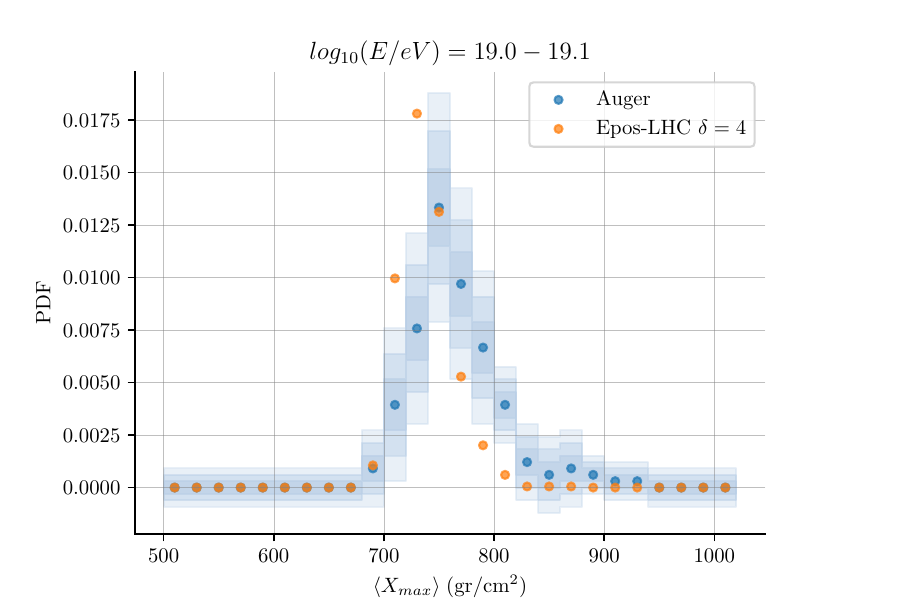} }}%
    {{\includegraphics[width=0.5\textwidth]{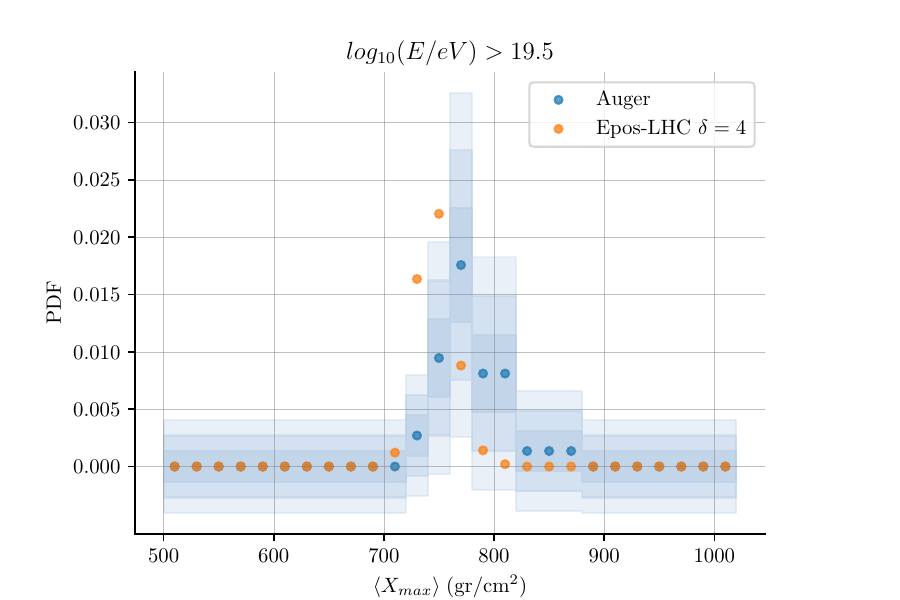} }}%
    \caption{$X_{max}$ distribution within the EPOS-LHC framework, compared against the case with $\delta=4$, for three energy bins, spread out across the spectral range investigated here. Auger data are taken from from \cite{DistPaper2017}.}%
    \label{fig:7}%
\end{figure}

\begin{figure}%
    \centering
    [\centering QGSJETII-04] {{\includegraphics[width=0.5\textwidth]{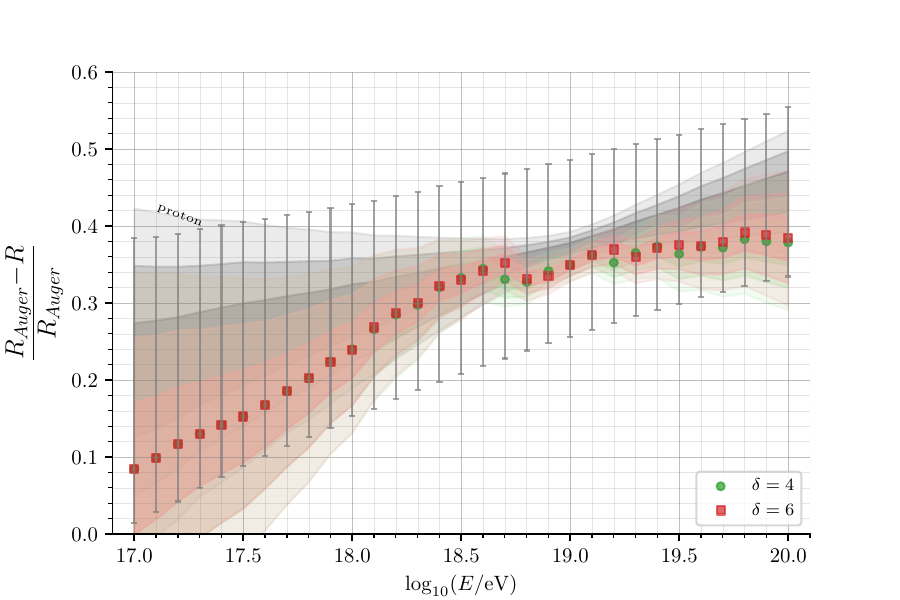} }}%
    \hspace{-10pt}
    [\centering EPOS LHC] {{\includegraphics[width=0.5\textwidth]{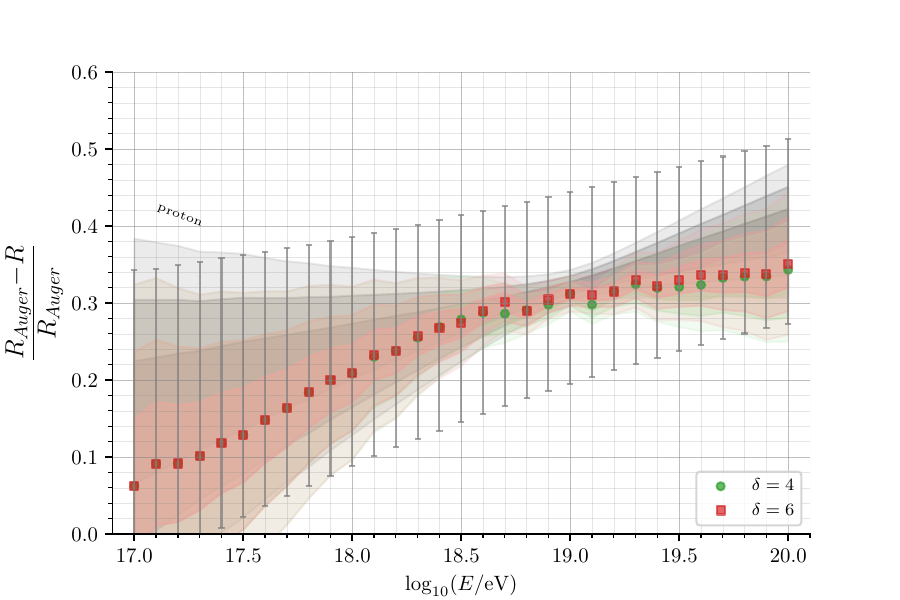} }}%
    \caption{Fractional difference of muons on the ground predicted by simulations relative to the observed data at the Auger Observatory  (data taken from \cite{PierreAuger:2021qsd})}. When new physics sets in above $E_\text{th}$, the number of muons change (green and red), due to the change in product multiplicity (hence, independently of $\delta$). This change does not fully reconcile the simulated number of muons with observations. However, unlike the SM prediction (grey), the discrepancy does not increase with energy above $10^{18.5}$ eV, but rather stabilizes around $30\%$ ($40\%$) for EPOS-LHC (QGSJETII-04).%
    \label{fig:5}%
\end{figure}

\begin{figure}%
    \centering
    [\centering QGSJETII-04] {{\includegraphics[width=0.5\textwidth]{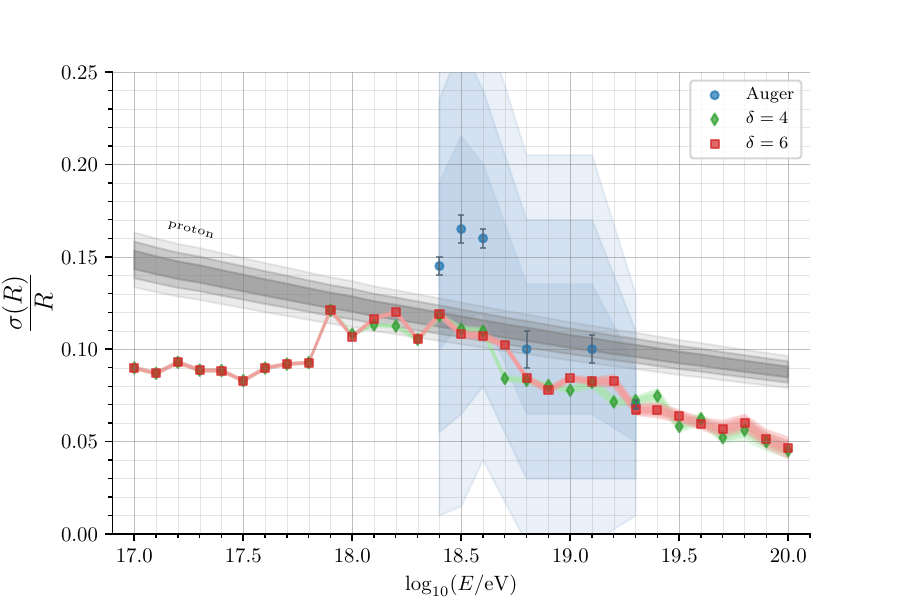} }  }%
    \hspace{-10pt}
    [\centering EPOS LHC] {{\includegraphics[width=0.5\textwidth]{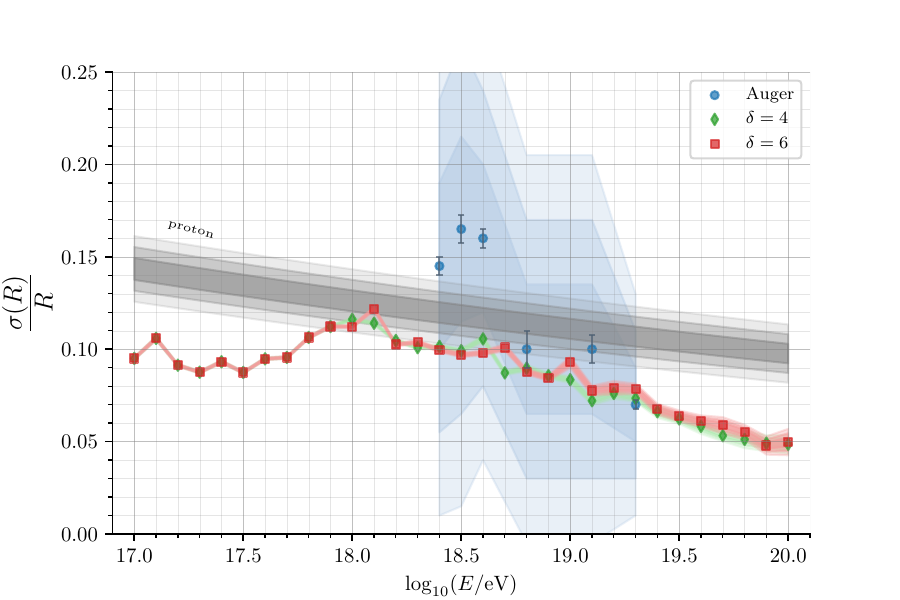} }}%
    \caption{Standard deviation over average of the muon ratio  as a function of energy. When new physics sets in above $E_\text{th}$, this quantity becomes more consistent with observed data than SM predictions (grey), and reproduces the trend with energy observed by Auger.}%
    \label{fig:6}%
\end{figure}

We next explore the agreement between new-physics $\sigma(X_{\text{max}})$ and observed Auger data, quantifying it by means of the reduced $\chi^2$ statistic. Since we have treated the Galactic component in a very approximate way, we only use energies at which extragalactic protons have fully dominated the UHECR flux ($E>10^{18.5}$ eV) for this comparison. In Fig.(\ref{fig:2}) we plot the reduced $\chi^2$  as a function of $\delta$ for our EAS simulation results (datapoints), as well as for our analytic approximation for $\sigma_{X_\text{max}}$ given by Eq.~(\ref{varxmaxnew}) (solid lines). We calculate the position of minimum $\chi^2$ for our simulations by fitting parabolas (dashed lines) to our datapoints. EPOS-LHC has a minimum of $\chi^2=1.7$ at $\delta=4.5$ and QGSJETII-04 has a minimum of $\chi^2=3.0$ at $\delta=6.7$. EPOS-LHC has overall better performance. At their minimum $\chi^2$ the cross section at an energy of $10^{19}$ eV rises to $802$ mb ($788$ mb) for EPOS-LHC (QGSJETII-04). Furthermore at the same energy the multiplicity has increased by a factor of $3.2$ ($1.8$).

 So far we have compared the new Physics scenario with Auger data using moments (mean and variance) of the $X_\text{max}$ distribution. It is however informative to also consider how results from the new Physics scenario compare with the full $X_\text{max}$ distribution. We present this comparison in  Fig.\ref{fig:7}, for the case of EPOS-LHC simulations with $\delta =4$, for three energy bins spread across the spectral range considered here, using Auger data from (\citep{DistPaper2017}. The results are in similarly good agreement as expected from the corresponding moments, with some discrepancy seen in the bin centered on $10^{17.85}$ eV, as is also the case for $\langle  X_\text{max}\rangle$.

In Figs. \ref{fig:5} and \ref{fig:6} we show how the changes in cross-section and multiplicity in our scenario affect the number of muons measured on the ground. Fig. \ref{fig:5} shows the fractional difference of muons on the ground from simulations relative to the data observed by Auger  \cite{PierreAuger:2021qsd}. As in all figures, bands represent statistical uncertainties and error bars systematic uncertainties, entering through the Auger measurements, and propagated to the plotted quantity. The new-physics scenario produces more muons than the SM prediction.  Although this is an improvement, still a deviation between $30-37\%$ from observational data persists. 
The reason is that the muon production does not depend strongly on the overall multiplicity of the secondary particles but rather on the ratio of pions and kaons produced after the first interaction \citep{Albrecht:2021yla}. The fraction of such particles in the products of the first collision does not change in the implementation of new physics we have considered here. However, such a change in the charged-particle ratio may be an important feature in any specific new-physics model that attempts to fully explain the UHECR composition problem, including the muon problem. We do however point out that the residual discrepancy between simulated and observed number of muons in our current implementation is constant with energy, unlike the SM predictions, where the discrepancy increases with energy. 
Our implementation additionally produces a {\em variance} for the muon number that is consistent with observed data (Fig.\ref{fig:6}), including the correct trend with energy. We plan to revisit this issue in a future publication, adding the possibility of a change in the charged-particle ratio.
   
\section{Summary and Conclusions}\label{sec6}

We performed simulations of EAS with CORSIKA, appropriately modified for interactions above $E_\text{th}=10^{18} \text{ eV}$ to feature an increased proton-air cross-section and first-collision product multiplicity. Ww have parameterized the increase in cross section through a parameter $\delta$, defined as the fractional increase of the coefficient of logarithmic growth with energy of the proton-air cross-section with respect to its standard-model value (see Eq.~\ref{eq:sigma_new}). We have parameterized the increase in product multiplicity through a function $n(\varepsilon)$, defined as the ratio of first-collision products over their SM-predicted number. By demanding that Auger observations of $\langle X_\text{max} \rangle$ as a function of energy above $10^{18.7}$ eV are reproduced under the assumption that all UHECR at these energies are protons, we can determine $n(\varepsilon)$ for any given value of $\delta$ (see Eq.~\ref{multiplicity}). This then leaves $\delta$ as the only parameter in our description. 

We have shown that these modifications to hadronic interactions at energies above $10^{18}$ eV are sufficient to reproduce observations of $X_\text{max}$. The growth of  $\langle X_\text{max} \rangle$ is reproduced for any value of $\delta$ (by construction), and $\sigma_{X_\text{max}}$ is best reproduced for $\delta$ between 4 and 7. If QGSJETII-04 is used for SM modeling of high-energy hadronic interactions, the optimal value for $\delta$ is 6.7. For EPOS-LHC, the optimal value for $\delta$ is  4.5. Epos-LHC with  $\delta=4.5$ provides the best overall fit to Auger $X_\text{max}$ data at $E>10^{18.5}$ eV. In each case, product multiplicity increases by a factor $n(\varepsilon)$ given by Eq.~(\ref{multiplicity}). 

These results provide phenomenological constraints on the properties (cross section, multiplicity) of any new effect beyond the SM that may set it for collisions at CM energies exceeding $\sim 50$ TeV, if such an effect is to be held responsible for the change of behavior of the $X_\text{max}$ distribution observed by Auger for collisions of primaries more energetic than $10^{18.7}$ eV with the atmosphere. 

 Consistently with expectations from the analytic estimates in PT19, modeling the increase in multiplicity as a set of "component showers" developing independently leads to a decrease in $X_{\rm max}$ variance, due to the repeated sampling from the parent distribution of shower products. We have thus confirmed in shower simulations the deviation of our results from those of \cite{Ulrich:2010rg}, who used a multiplicative factor 
to increase the number of products in each collision. In that work, a multiplicative increase of products of identical distribution as the original shower left $\sigma_{X_\text{long}}$ unchanged. We conclude that the discrepant results between the two works are due to the differences in fundamental assumptions about how a multiplicity increase would manifest phenomenologically, rather than some numerical issue.

As far as the muon problem is concerned, we found that although the change in multiplicity we have investigated here does improve the agreement between Auger data and EAS simulations, it is not by itself sufficient to fully resolve the discrepancy.  The source of the muon observables tension with simulations cannot be traced to multiplicity and cross section effects.  It is conceivable that a new effect setting in at 50 TeV might also induce a change in the charged-particle ratio, which would further increase the number of muons produced and detected on the ground. We plan to investigate this possibility in a future publication.  

Current and planned advances in the tomographic mapping of the Galactic magnetic field through local measurements \citep{PASIPHAEwhite,OtherTritsis,2019ApJ87338T, ClarkHensley2019, Skalidis1, Skalidis2} are expected to make feasible an electromagnetic determination of the charge of UHECR of the highest energies in the near future \citep{MP2019}. If such studies provide unequivocal evidence that UHECR at the highest energies are indeed protons, then this will be a strong argument in favor of new physics setting in for hadronic interactions at CM energies above 50 TeV, with the phenomenology of any such new effect exhibiting the behavior calculated in this work. 

\begin{acknowledgments}
S.R. and V.P. would like to dedicate this work to the memory of our friend and collaborator Theodore Tomaras whom we have lost way too soon. We will always fondly remember all the exciting and fruitful scientific discussions and debates that we have had and we will be sorely missing the ones that will now never take place. We thank Alan Watson, Nicusor Arsene, Konstantina Dolapsaki, Andreas Tersenov, and Christos Litos for helpful comments and discussions that improved this manuscript, and Dieter Heck for valuable feedback on the use of CORSIKA. This work was supported  by the Hellenic Foundation for Research and Innovation (H.F.R.I.) under the “First Call for H.F.R.I. Research Projects to support Faculty members and Researchers and the procurement of high-cost research equipment grant” (Project 1552 CIRCE). V.P. acknowledges support from the Foundation of Research and Technology - Hellas Synergy Grants Program through project MagMASim, jointly implemented by the Institute of Astrophysics and the Institute of Applied and Computational Mathematics.

\end{acknowledgments}

\bibliography{PRD_New_Physics}

\end{document}